\documentclass{aa}
\usepackage{epsfig,graphicx}
\begin{document}
\title{A non-LTE analysis of the spectra of two narrow lined main sequence
stars in the SMC}


\author{I. Hunter \inst{1}
\and
P.L. Dufton\inst{1}
\and 
R.S.I. Ryans\inst{1}
\and
D.J. Lennon\inst{2}
\and
W.R.J. Rolleston\inst{1}
\and
I. Hubeny\inst{3}
\and
T. Lanz\inst{4}}

\institute{APS Division, Department of Pure \& Applied Physics, The Queen's University 
           of Belfast, BT7 1NN, Northern Ireland, UK 
     \and
       Isaac Newton Group of Telescopes, Apartado de Correos 368, E-38700, 
       Santa Cruz de La Palma, Canary Islands, Spain
     \and
       Steward Observatory, University of Arizona, Tucson, AZ\,85712, USA
     \and
       Department of Astronomy, University of Maryland, College Park, MD\,20742,
       USA}
       
\offprints{I. Hunter,\\ \email{I.Hunter@qub.ac.uk}}

\date{Received 2 November 2004; accepted 20 February 2005}

\abstract{An analysis of  high-resolution VLT/UVES spectra of two B-type 
main sequence stars, NGC\,346-11 and AV\,304, in the Small Magellanic Cloud 
(SMC), has been undertaken, using the non-LTE {\sc tlusty} model atmospheres 
to derive the stellar parameters and chemical compositions of each star.
The chemical compositions of the two stars are in reasonable agreement. Moreover, our stellar
analysis agrees well with earlier analyses of \ion{H}{ii} regions. The results
derived here should be representative of the current base-line chemical
composition of the SMC interstellar medium as derived from B-type stars.
    \keywords{stars: abundances -- stars: atmospheres -- 
             stars: early-type -- Galaxies:individual: Small Magellanic Cloud
               }
   }
\titlerunning{Non-LTE analysis of main sequence stars in the SMC}

\maketitle{}
%
\section{Introduction}                                         \label{s_intro}

The Magellanic Clouds have been intensively studied in terms of, for example, 
their star formation history (see, for example, Yoshizawa \& Noguchi \cite{yos03}; 
Harris \& Zaritsky \cite{har04}, Zaritsky et al. \cite{zar04})), kinematics 
(see, for example, Stanimirovic et al. \cite{sta04}; Kim et al. \cite{kim03}; 
Gardiner \& Noguchi \cite{gar96}), stellar populations (see, for example, 
Cioni \& Habing \cite{cio04}; Evans et al. \cite{eva04}; Kunkel et al. \cite{kun95})
and chemical compositions (see, for example, the reviews of Garnett \cite{gar99}
and Westerlund \cite{wes97}). This interest arises from their relative
proximity (Harries et al. \cite{har03}), which means that it is possible
to study individual stars and nebulae in detail. Additionally
their relatively low extinction allows them to be viewed in their entirety.
Their very different environments to that of our own Galaxy then makes
them ideal for studying both stellar and galactic evolution.

In the case of mapping the chemical compositions of the Magellanic Clouds, 
a variety of targets
have been studied including \ion{H}{ii} regions (see, for example, Russell \&
Dopita \cite{rus92}, Garnett \cite{gar99}, Kurt et al. \cite{kur99}), 
the interstellar medium, ISM (Welty et al. \cite{wel97, wel99}), 
early-type (Bouret et al. \cite{bou03}; Korn et al. \cite{kor00, kor02}; 
Rolleston et al. \cite{rol03}; Trundle et al. \cite{tru04}) and late-type
(Barbuy et al. \cite{bar91}; Spite et al. \cite{spi89, spi91}; Hill 
\cite{hil97,hil99}) stars. All these methods have particular strengths 
and limitations that can be characterised under two main headings. Firstly 
there is the physical model used in predicting the observed spectra. This 
model will make assumptions about the geometry of and physical conditions 
in the plasma and about the transfer of radiation, which will impact on the
reliability of the derived chemical compositions. Equally important is
the nature of the plasma that is observed. For example, ISM studies 
normally observe only the gas phase component, whilst the chemical
compositions of stellar atmospheres may have been modified by mixing
of nucleosynthetic material to the surface. The latter is particularly
important as even relative unevolved B-type giants (Lennon et al.
\cite{len03}; Korn et al. \cite{kor00}) and O-type dwarfs 
(Bouret et al. \cite{bou03}) appear to have contaminated atmospheres.

Such effects can be used to understand the physical processes
that are occuring, e.g. the chemical composition and nature of ISM grains
and the detail of how stars evolve. However these limit the usefulness
of such techniques in estimating the current chemical composition of the
interstellar medium in the host galaxy. Historically \ion{H}{ii} regions have
been used for such studies as they directly sample the ISM and their emission
line spectra are relatively easy to observe. Recent SMC studies  show relatively
good agreeement (see, for example,  Russell \& Dopita \cite{rus92}, Reyes
\cite{rey99}, Garnett \cite{gar99}, Kurt et al. \cite{kur99}, Peimbert et al.
\cite{pei00} and Testor \cite{tes01}), illustrating the utility of the
method and indirectly implying that the SMC is relatively well mixed. However,
it is important that other methods are available both to investigate the
possibility of systematic errors and also to extend the range of elements that
can be studied.  The spectra of main sequence B-type stars offer such an
alternative as their atmospheres should be uncontaminated and indeed Korn et al.
(\cite{kor02}) have studied such objects in the LMC. Additionally two
narrow lined B-type stars, AV\,304 and NGC\,346-11 have been identified 
in the SMC and analysed by Dufton et al. (\cite{duf90}) and Rolleston et al.
(\cite{rol93, rol03}). These analyses were limited by their adoption of an LTE
model atmosphere approach and for the two earlier analyses, by the moderate
quality of the spectroscopic data for these relatively faint targets.
Here we present non-LTE analyses of these two targets based on high
quality UVES/VLT observations that should complement the 
existing \ion{H}{ii} region SMC studies. 

\section{Observations and data reductions}             \label{s_obs}

\begin{figure*}
\centering
\epsfig{file=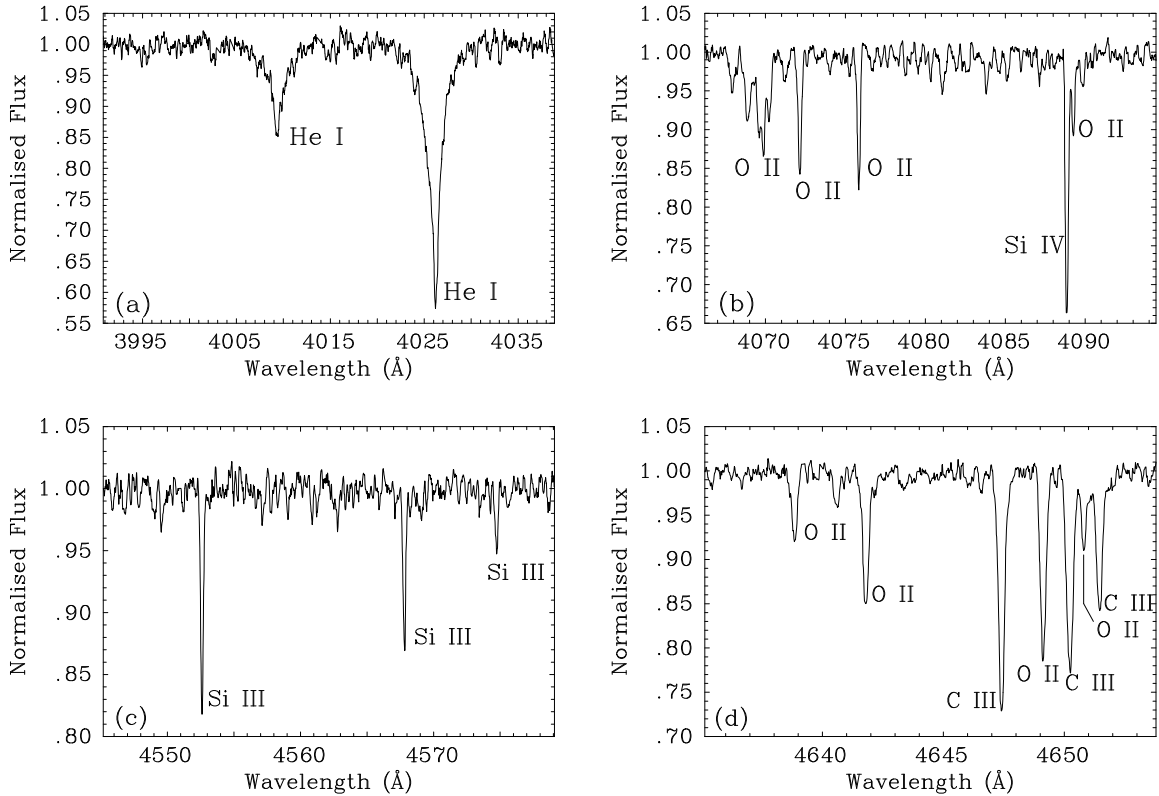,height=160mm, angle=-90}
\caption[]{Examples of the observed spectra of NGC\,346-11 from UVES. These
sample spectral regions include important lines such as \ion{Si}{iv} (b),
\ion{Si}{iii} (c) and \ion{C}{iii} (d). Notice the lack of an obvious
\ion{N}{ii} line at 3995\AA~in (a).}
\label{f_lines}
\end{figure*}

High-resolution spectra have been obtained for both NGC\,346-11 and AV\,304
using the Ultraviolet and Visual Echelle Spectrograph, UVES, (D'Odorico et al.
\cite{dod00}) on the UT2 (Kueyen) telescope at the European Southern
Observatory. NGC\,346-11 was observed during a three night run in November 2001,
whilst the observations of AV\,304, taken during a two night run in January
2001, have been previously discussed by Rolleston et al. (\cite{rol03}). UVES
was operated using a two arm cross-disperser with CCD-44 chips, with slit widths
of 1.5 arcsec and 1.0 arcsec for NGC\,346-11 and AV\,304 respectively. Complete
spectral coverage in the range 3770-4980\AA\ was obtained for both stars and the
echellograms were reduced to a one dimensional format using UVES pipe-line
software. Combining the individual exposures resulted in signal-to-noise ratios 
of approximately 100 and 80 for NGC\,346-11 and AV\,304 respectively. Examples
of sample regions of the observed spectra for NGC\,346-11 and AV\,304 are displayed in
Fig.~\ref{f_lines} and Rolleston et al. (\cite{rol03}) respectively. These
spectral regions clearly show the high resolution and signal-to-noise ratio of
our data.

For the NGC\,346-11 spectrum, data reduction followed the procedures discussed
by Rolleston et al. (\cite{rol03}). The spectrum analysis package {\sc dipso}
(Howarth et al. \cite{how94}) was used to normalise and then radial 
velocity correct the spectra. Equivalent widths were measured for
all the observable absorption metal lines by non-linear least squares fitting 
using Gaussian profiles and a low order polynomial to represent the
continuum. For well observed isolated features, an error estimate of 
normally less than 10\% was found for the equivalent width measurements, 
whilst this increased to typically 20\% for blended or weak absorption lines.
The equivalent width estimates for both targets are listed in Table~1 available
at the CDS. Table~1 contains the following information; column 1 lists the
species of the observed spectral line, column 2 lists the rest wavelength of the
observed line, column 3 lists the equivalent width of the observed lines in
AV\,304, column 4 lists the abundances derived from the equivalent widths and
atmospheric paramters of AV\,304, column 5 lists the equivalent width of the
observed lines in NGC\,346-11 and column 6 lists the abundances derived from the
equivalent widths and atmospheric paramters of NGC\,346-11.

\section{Analysis}                              \label{s_atmos}

\subsection{Non-LTE atmosphere calculations}           \label{s_nlte}
The analysis is based on grids of non-LTE model atmospheres calculated 
using the codes {\sc tlusty} and {\sc synspec} (Hubeny \cite{hub88}; 
Hubeny \& Lanz \cite{hub95}; Hubeny et 
al.\, \cite{hub98}). Details of the methods can be found in Ryans et al.\ 
(\cite{rya03}), while the grids have been discussed in more detail by Dufton et 
al.\ (\cite{duf05}) and hence we will limit ourselves to a short overview of
the methods.

Briefly four grids have been generated with base metallicities corresponding 
to our Galaxy ($[\frac{Fe}{H}]$ = 7.5\,dex) and with metallicities reduced by
0.3, 0.6 and 1.1\,dex. These lower metallicities were chosen so as to be
representative of the LMC, SMC and 
low metallicity material. For each base metallicity,
approximately 3\,000 models have been calculated covering a range of effective
temperature from 12\,000 to 35\,000\,K, logarithmic gravities (in cm s$^{-2}$)
from 4.5\,dex down to close to the Eddington limit (which will depend on the
effective temperature) and microturbulences of 0, 5, 10, 20 and 30\,km~s$^{-1}$.
Then for any set of atmospheric parameters, five full independent models were 
calculated keeping the iron abundance
fixed but allowing the abundances of the light elements, C, N, O, Mg, Si and S, 
to vary from +0.8\,dex
to $-$0.8\,dex around their base values. Effectively this approaches assumes
that the line blanketing and atmospheric structure is dominated by iron and
hence that the light element abundances can be varied without significantly
affecting this structure. Tests discussed in Dufton et al.\,(\cite{duf05})
appear to confirm that this approach is reasonable. 

These models are then used to calculate spectra, which in turn provide
theoretical hydrogen and helium line profiles and equivalent widths for light
metals for a range of abundances. The
theoretical equivalent widths are then available via a GUI interface written
in IDL, which allows the user to interpolate in order to calculate equivalent
widths and/or abundance estimates for approximately 200 metal lines for any
given set of atmospheric parameters. Ryans et al. (\cite{rya03}) reported that
the increments of 0.4\,dex used in our grids were fine enough to ensure that no
significant errors were introduced by the interpolation procedures. Full
theoretical spectra are also available for any given model. Details of the
model atmosphere grids, atomic data used in the line strength calculations and wavelength
ranges used in the equivalent width calculations are discussed in Dufton et al.
(\cite{duf05}) and details are also avaliable at http://star.pst.qub.ac.uk/. 

\subsection{Stellar Atmospheric Parameters}                      \label{atm_par}

Four parameters define the basic
characteristics of the stellar atmosphere - effective temperature ($T_{\rm
eff}$), logarithmic gravity ($\log g$), mircoturbulence ($\xi$), and the
metallicity (iron content) of the star. As these parameters are all 
inter-related, one must use an iterative process but by careful 
selection of the initial estimates, in most cases 
only two or three iterations were necessary. Standard techniques were used 
and as they have previously been 
discussed by, for example, Kilian (\cite{kil92}), Kilian et al. (\cite{kil94}), 
McErlean et al. (\cite{mce99}), Korn et al. ({\cite{kor00}) and Trundle et al. 
(\cite{tru04}), they will only be briefly discussed here. Note that 
the grid of models with an iron abundance
of 6.9\,dex was initially used to determine the other parameters and the
effect of varying this value will be discussed below.

\subsubsection{Effective Temperature, $T_{\rm eff}$}            \label{s_teff}

The \ion{Si}{iii} to \ion{Si}{iv} ionization
equilibrium was used for estimating the effective temperature and the
adopted values are listed in Table~\ref{t_atmos+abund}. Using the grid
of models with an iron abundance of 6.4\,dex changed these estimates by less
than 500\,K and hence the adopted model metallicity is unlikely to
be a serious source of error. The errors due to uncertainties in the 
observational data are estimated to be about $\pm$1000\,K, whilst those
due to the adopted atomic data and physical assumptions
are difficult to quantify. However Dufton et al. (\cite{duf05}) discussed the
the analysis of two SMC B-type supergiants using 
the current grids and the unified code {\sc fastwind} 
(Santolaya-Rey et al. \cite{san97})
and found encouraging agreement. Additionally for NGC\,346-11, the
\ion{He}{ii} spectrum could be observed and provided an independent 
estimator. This implied an effective temperature of approximately 33\,500K,
which is consistent within the uncertainties in the temperature deduced from the 
silicon lines (see Fig. \ref{f_He}).

The only other element in which we see two ionization stages is carbon and hence
the ionization equilibrium of \ion{C}{ii} to \ion{C}{iii} may also be used to
deduce $T_{\rm eff}$. From Table~\ref{t_atmos+abund} it can be seen that the
carbon abundances 
derived from each species are in excellent agreement for NGC\,346-11 but the
abundances differ by 0.3\,dex for AV\,304. Using the \ion{C}{ii} to
\ion{C}{iii} ionization equilibrium to derive the effective temperature for
AV\,304 would result in a temperature of 28300\,K which is within the uncertainity
and hence consistent with the \ion{Si}{iii} to \ion{Si}{iv} ionization equilibrium
temperature.

\begin{table*}
\addtocounter{table}{+1}
\caption[]{The stellar parameters and absolute abundance estimates, together
with their estimated uncertainties, for NGC\,346-11 and AV\,304. The 
quantities inside brackets are the number of lines used to estimate
the abundances. For the ions denoted with a $^{*}$ the abundance estimates have 
been calculated using LTE rather than non-LTE methods as discussed in 
Sect.~\ref{s_LTE}.}
\label{t_atmos+abund}
\centering
\begin{tabular}{lllllll} \hline \hline
 & \multicolumn{3}{c}{NGC\,346-11} & \multicolumn{3}{c}{AV\,304} \\ \hline 
$T_{\rm eff}$(K)   & 32\,500 & $\pm$1000 &     & 27\,500 & $\pm$1000 &    \\
$\log g$(dex)      & 4.25    & $\pm$0.20 &     & 3.90    & $\pm$0.20 &    \\
$\xi$(km~s$^{-1}$) & 5       & $\pm$5    &     & 3       & $\pm$3    &    \\
Spectral            & B0V     &           &     & B0.5V   &           &    \\
Type                &         &           &     &         &           &     \\
\\
\ion{C}{ii}         & 7.45    & $\pm$0.29 & (2) & 7.36    & $\pm$0.12 & (3)\\
\ion{C}{iii}        & 7.44    & $\pm$0.18 & (2) & 7.66    & $\pm$0.30 & (2)\\
\ion{N}{ii}         & 6.73    & $\pm$0.31 & (1) & 6.55    & $\pm$0.18 & (1)\\
\ion{O}{ii}         & 7.82    & $\pm$0.20 & (22)& 8.13    & $\pm$0.10 & (42)\\
\ion{Mg}{ii}        & 6.77    & $\pm$0.23 & (1) & 6.77    & $\pm$0.16 & (1)\\
\ion{Si}{iii}       & 6.42    & $\pm$0.25 & (3) & 6.76    & $\pm$0.19 & (4)\\
\ion{Si}{iv}        & 6.39    & $\pm$0.27 & (3) & 6.73    & $\pm$0.44 & (2)\\ 
\ion{Ne}{ii}$^{*}$  & -	      & -         &	& 7.84    & $\pm$0.16 & (2)\\
\ion{Al}{iii}$^{*}$ & -       & -         &     & 5.33    & $\pm$0.11 & (1)\\
\ion{S}{iii}        & -       & -         &     & 6.40    & $\pm$0.15 & (7)\\
\ion{Fe}{iii}$^{*}$ & -       & -         &     & 6.63    & $\pm$0.23 & (4)\\
\hline
\end{tabular}
\begin{itemize}
\footnotesize
\item[] The spectral types of NGC\,346-11 and AV\,304 have been obtained from
Rolleston et al. (\cite{rol93}) and the {\sc simbad} database, operated at CDS,
Strasbourg, France respectively.
\normalsize
\end{itemize}
\end{table*}

\begin{figure}
\centering
\epsfig{file=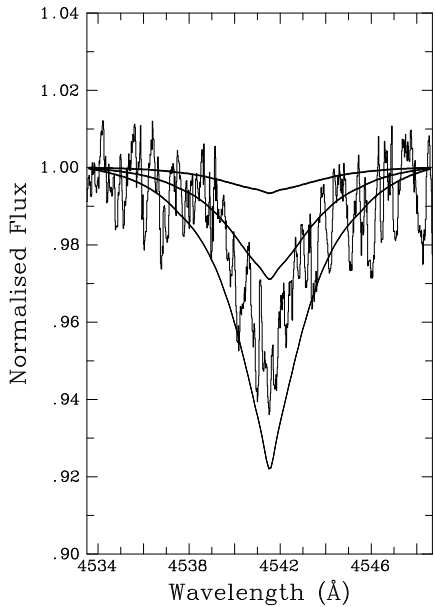,height=90mm, angle=-0}
\caption[]{Observed and theoretical spectra for the \ion{He}{ii} line at 4541\AA~in NGC\,346-11.
The latter are for effective temperatures of 30000K, 32500K and 35000K 
(upper, middle and lower smooth curves respectively) and imply an estimate
of approximately 33\,500K.}
\label{f_He}
\end{figure}
 
\subsubsection{Logarithmic Surface Gravity, $\log g$}          \label{s_logg}

Surface gravity estimates were determined by fitting the observed Balmer series 
lines (H$\beta$, H$\gamma$ and H$\delta$) with theoretical profiles, with
agreement between the different estimates being excellent.
The adopted values are listed in Table~\ref{t_atmos+abund}.
Observational and fitting errors imply an uncertainty of 
$\pm$0.1-0.2\,dex, whilst an error of 1000\,K in the effective temperature
would introduce an additional uncertainty of approximately 0.1\,dex.
As for the estimation of the effective temperatures, the use of the 
grid with an iron abundance of 6.4\,dex yielded similar gravity estimates.
The quality of the agreement between observation and theory is 
illustrated in Fig. \ref{f_fits} for both NGC\,346-11 and AV\,304.
The \ion{He}{ii} line presented in Fig. \ref{f_He} can also be used to estimate 
$\log g$ for NGC\,346-11. This line implied a value of 4.1\,dex which is well within 
our uncertainity and given the sensitivity of this measurement to our adopted
effective temperature we have no reason to believe that this value is more 
appropiate than that given in Table~\ref{t_atmos+abund}.

\begin{figure}
\centering
\epsfig{file=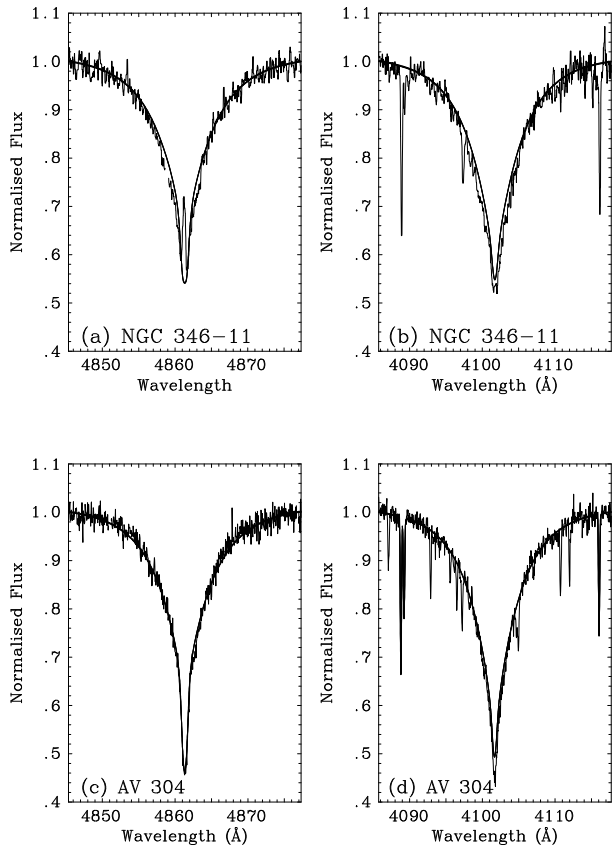, height=120mm, angle=-0}
\caption[]{Examples of the agreement between observed and theoretical
(generated using the atmospheric parameters listed in
Table~\ref{t_atmos+abund}) Balmer line profiles.}
\label{f_fits}
\end{figure}

\subsubsection{Microturbulence, $\xi$}                      \label{s_xi}

The microturbulence, as in other studies (see for example
Vrancken et al. \cite{vra00} and Trundle et al. \cite{tru04}), proved to be a
difficult quantity to determine accurately. The standard technique of 
eliminating the dependence of the estimated abundance on the line strength 
(see for example Korn et al. \cite{kor00}) was used for the \ion{O}{ii} 
or \ion{Si}{iii} spectra. The large
number of \ion{O}{ii} spectral lines should lead to robust estimates of $\xi$,
but the validity of this method may be compromised by the \ion{O}{ii} lines 
arising from different multiplets. For the \ion{Si}{iii} multiplet at 
4560\AA, this complication is removed but we are now reduced to only 
using three lines for the evaluation of $\xi$.  

From the \ion{O}{ii} lines in AV\,304, a value of $\xi$ of 3\,km~s$^{-1}$ was 
estimated. By contrast a microturbulence of 1\,km~s$^{-1}$ was implied by the
\ion{Si}{iii} multiplet. We have adopted a microturbulence in AV\,304 of
3$\pm$3\,km~s$^{-1}$ to allow for the uncertainty in this parameter. It should
be noted that this uncertainty does not have a significant effect on the other
atmospheric parameters.

NGC\,346-11 has a higher effective temperature and the metal absorption
lines are generally weaker. Indeed for the \ion{O}{ii} lines, the small range 
of equivalent widths precluded a reliable estimate of the microturbulence.
The \ion{Si}{iii} lines yielded an estimate for $\xi$ of 5\,km~s$^{-1}$ but 
it is not possible to rule out higher values and an uncertainty of 
$\pm$5\,km~s$^{-1}$ was therefore adopted. Fortunately the weakness of the 
metal line spectrum in this star leads to the abundance estimates being 
relatively insensitive to this quantity and again this uncertainty does not have
a significant effect on the other atmospheric parameters.

\subsubsection{Projected Rotational Velocity, $v\sin i$}       \label{s_vsini}

Both NGC\,346-11 and AV\,304 are known to have very low projected rotational
velocities since the metal lines appear sharp in their spectra. 
Rotational velocities can be estimated for each star by rotationally broadening 
the theoretical spectra, using the procedures discussed in Gray (\cite{gra92}),
until they match the observed spectra. Six strong, well defined oxygen and 
silicon lines were considered, with the theoretical spectra
being taken from our non-LTE grid. A projected rotational velocity,
$v\sin i$, of 8.0$\pm$2.0 km~s$^{-1}$  
was deduced for NGC\,346-11 with the uncertainty 
being the 1\,$\sigma$ standard deviation of the individual estimates. The
theoretical spectra for AV\,304 at a microturbulence of 3~km~s$^{-1}$ is
of a similar width to that of the observed spectra and as such any estimate 
will be highly dependent on the mircoturbulence adopted. 
Hence, it is not possible to reliably estimate $v\sin i$
although it is clear that it is very low and the observed spectra is
dominated by the instrumental resolution. The very
low projected rotational velocities make
NGC\,346-11 and AV\,304 ideal for estimating their
chemical compositions. However if they indeed have small rotational 
velocities rather than small angles of inclination, it is possible that they
are not representative of the (near) main sequence hot star populations
of the SMC.

\subsection{Non-LTE photospheric abundances}                    \label{s_abund}

Non-LTE photospheric abundance estimates were derived for NGC\,346-11 
and AV\,304 using the stellar parameters discussed in Sect. \ref{atm_par} 
and are listed in Table~\ref{t_atmos+abund}. The standard logarthmic scale for
presenting abundance estimates has been adopted, where hydrogen is taken 
to have an abundance of 12.0 dex and all the abundances are relative to this 
value. The abundances derived from each line in both stars can be found in Table~1.

The uncertainties listed for the photospheric abundances in
Table~\ref{t_atmos+abund} are calculated in two parts. The first considers the
random errors associated with analysing the data, e.g. observational
uncertainties, individual errors in oscillator strengths etc.
This error is taken to be the standard deviation of
the derived abundances in each species divided by the square root of the number of
absorption lines observed. In cases where only one line of a species was
observed, random
errors were considered to be equal to the standard deviation of the most observed
species (\ion{O}{ii}) but given that in most cases the systematic errors
are larger than the random errors, this assumption should not be critical. The second
part considers systematic errors arising from the stellar atmospheric parameter
estimates. All the stellar  parameters were increased in turn by their
uncertainty and the mean abundance derived was
compared to the absolute abundance listed in Table~\ref{t_atmos+abund}. In
Table~\ref{t_errors} we have listed the uncertainties in our abundance estimates due to both the
uncertainty arising from random errors and to the uncertainties in the adopted atmospheric
parameters.

\begin{table*}
\caption[]{A breakdown of the uncertainities in the abundance of each species due to both the
uncertainity arising from random errors and the uncertainity in each of the adopted atmospheric
parameters.}
\label{t_errors}
\centering
\begin{tabular}{lllllllll} \hline \hline
Species & \multicolumn{4}{c}{NGC\,346-11} & \multicolumn{4}{c}{AV\,304} \\ 
             &$\sigma_{\rm obs}$ & $\sigma_{\rm T}$ & $\sigma_{\rm g}$ & $\sigma_{\rm \xi}$ & $\sigma_{\rm obs}$ & $\sigma_{\rm T}$ & $\sigma_{\rm
	     g}$ & $\sigma_{\rm \xi}$  \\ \hline
\\
\ion{C}{ii}         &0.17&0.21&0.11&0.03&0.05&0.10&0.04&0.00\\
\ion{C}{iii}        &0.04&0.10&0.09&0.11&0.01&0.23&0.18&0.06\\
\ion{N}{ii}         &0.22&0.19&0.09&0.04&0.14&0.11&0.04&0.01\\
\ion{O}{ii}         &0.05&0.17&0.09&0.01&0.02&0.06&0.02&0.07\\
\ion{Mg}{ii}        &0.22&0.05&0.01&0.04&0.14&0.06&0.02&0.04\\
\ion{Si}{iii}       &0.01&0.20&0.04&0.15&0.05&0.08&0.01&0.16\\
\ion{Si}{iv}        &0.12&0.10&0.15&0.16&0.09&0.28&0.26&0.19\\ 
\ion{Ne}{ii}        &   -&   -&   -&   -&0.05&0.12&0.09&0.03\\
\ion{Al}{iii}       &   -&   -&   -&   -&0.06&0.09&0.03&0.02\\
\ion{S}{iii}        &   -&   -&   -&   -&0.12&0.00&0.07&0.06\\
\ion{Fe}{iii}       &   -&   -&   -&   -&0.18&0.14&0.01&0.02\\
\hline
\end{tabular}
\begin{itemize}
\footnotesize
\item[] $\sigma_{\rm obs}$ is the standard deviation in the abundances of a given species divided by the number of observed lines; 
$\sigma_{\rm T}$ is the systematic uncertainity in the abundance given an uncertainity in $T_{\rm eff}$ of $\pm$1000\,K; 
$\sigma_{\rm g}$ is the systematic uncertainity in the abundance given an uncertainity in $\log g$ of $\pm$0.2\,dex; 
$\sigma_{\rm \xi}$ is the systematic uncertainity in the abundance given an uncertainity in $\xi$ of $\pm$5\,km~s$^{-1}$; 
Note that $\sigma_{\rm obs}$ for \ion{N}{ii} and \ion{Mg}{ii} is taken as the standard deviation of the \ion{O}{ii} abundances.
\normalsize
\end{itemize}
\end{table*}
The
total uncertainty in an abundance estimate for a given species was taken as the
square root of the sum of the squares of the uncertainty arising from 
the random errors and the systematic errors arising from the
temperature, gravity and microturbulence estimates. In the majority of cases
decreasing each of the stellar parameters in turn by their associated uncertainty leads to 
the same uncertainty as increasing each of the stellar parameters. In cases where there
was asymmetry, the uncertainty was taken as the average 
from increasing and decreasing each of the stellar parameters. 

\subsection{Other abundance estimates}			\label{s_LTE}

As discussed by Rolleston et al.\ (\cite{rol03}), lines arising from
\ion{Ne}{ii}, \ion{Al}{iii}, and \ion{Fe}{iii} were also observed
in the spectrum of AV\,304 and these species are not currently available
in our non-LTE grids. For completeness, we have calculated non-LTE models 
at the atmospheric parameters listed in Table~\ref{t_atmos+abund}. These
have then be used to derive abundance estimates (and uncertainties
using the methodology outlined in Sect. \ref{s_abund})
in an LTE approximation and these values are also listed in 
Table~\ref{t_atmos+abund}. Note that as we have used effectively the
same methods, there is no apriori reason to believe that these
estimates are superior to those deduced by Rolleston et al.\, (\cite{rol03}).

Although sulphur was included in our non-LTE calculations, no grids of
\ion{S}{iii} equivalent widths were calculated. However we have used the models
discussed above to generate theoretical non-LTE \ion{S}{iii} equivalent widths
and hence S abundance estimates. It is found that
our mean LTE abundance is only 0.01\,dex lower than the non-LTE value quoted in
Table~\ref{t_atmos+abund} and hence non-LTE effects appear to be negligible for
this species at the atmospheric parameters of AV\,304.

For other species, the main non-LTE effect may consist of overionization relative to LTE, but if we
use the lines of the dominant ion the LTE abundances will generally be in good
agreement with non-LTE abundances. At the atmospheric parameters of AV\,304
\ion{Fe}{iv} is the dominant species of iron and as such our LTE abundance
derived from the \ion{Fe}{iii} lines is likely to underestimate the actual iron
abundance in the SMC.

\subsection{Comparison with previous analyses}		\label{s_nlte_effect}

Rolleston et al. (\cite{rol03}) have previously presented an LTE analysis
for AV\,304 based on the same observational dataset. They deduced
similar atmospheric parameters with a slightly lower surface gravity
and higher microturbulence. However, within the estimated uncertainties
the two analyses are in agreement. The photospheric abundances are
also in reasonable agreement although there are differences ranging
up to 0.3\,dex. These are probably principally due to non-LTE effects and
Lennon et al. (\cite{len03}) discussed and tried to quantify these effects.
They report abundance estimates corrected for non-LTE effects of 7.41, 
6.55 and 8.16\,dex for the CNO elements respectively. These are in excellent agreement with our results  
for AV\,304 indicating that for these elements, the differences between our results and those of
Rolleston et al. (\cite{rol03}) arise from non-LTE effects. The remainder of the species for which we have used non-LTE methods
(Mg, Si and S) are in excellent agreement with the abundances reported by Rolleston et al.
(\cite{rol03}) and non-LTE effects are thought to be small for these species. For example, Trundle et al.
(\cite{tru04}) report non-LTE corrections for the Mg and Si abundances given in Rolleston et al.
(\cite{rol03}) to be approximately 0.05\,dex for both species.

Additionally Rolleston et al.\, (\cite{rol93}) undertook an LTE analysis of
lower quality AAT/IPCS spectroscopy for both AV\,304 and NGC\,346-11. 
They deduced relatively similar atmospheric parameters but with lower 
effective temperatures. The quality of their spectra limited the number
of ionic species that they could identify but for \ion{O}{ii} in AV\,304 and
\ion{C}{iii} in NGC\,346-11 agreement with the current analysis is good. 
For other species differences range from 0.2 to 0.4 dex and given the
improvements both in the quality of the observational data and in the
theoretical methods, we believe that the values presented here are more
reliable. 

\subsection{Interstellar lines}				\label{s_CaII}

A similar analysis to that discussed by Rolleston et al. (\cite{rol03}) for 
AV\,304 was used to estimate the LSR velocities of the individual 
components of the interstellar \ion{Ca}{ii} K line in NGC\,346-11. The velocity 
of the different components (rounded to the nearest km~s$^{-1}$) are tabulated 
in Table~\ref{t_CaII} and illustrated in Fig.~\ref{f_CaII}, with the results for
AV\,304 being taken directly from Rolleston et al. (\cite{rol03}). 

NGC\,346-11 shows fewer individual components, possibly due to the larger 
slit width (and hence lower spectral resolution) used in its observations.
Following similar methods to Rolleston et al. (\cite{rol03}) we associate
components N1 and N2, along with components A1 through to A4 with our Galaxy. 
We believe that the remaining components belong to the SMC. From the distinct
differences in the profiles, we may be observing
different SMC gas components although it is possible to identify interstellar 
components towards
the two stars with similar velocities (for example, 156-157 km~s$^{-1}$ and
200-201 km~s$^{-1}$). The targets have an angular separation of 0.53$^o$, 
corresponding
to approximately 500~pc for a distance modulus of 18.89
(Harries et al. \cite{har03}). Hence it is not clear whether these components
are related although Wayte (\cite{way90}) identified components that
had large spatial extents across both the SMC and LMC.

\begin{figure}
\centering
\epsfig{file=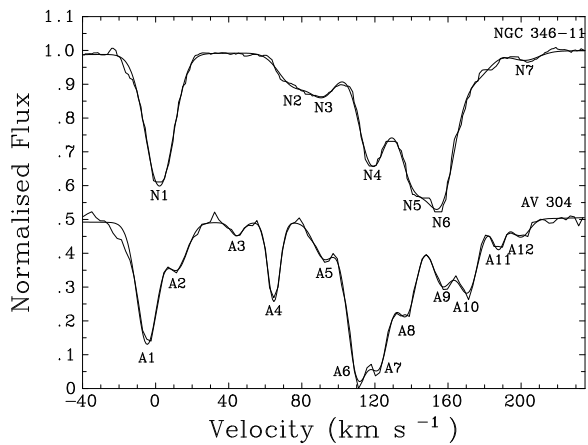, height=90mm, angle=-90}
\caption[]{The observed interstellar gas components towards NGC\,346-11 and
AV\,304, together with the Gaussian fits used to deduce the velocities 
listed in Table \ref{t_CaII}.}
\label{f_CaII}
\end{figure}

\begin{table}
\caption[]{LSR velocities (rounded to the nearest km~s$^{-1}$) of the different
components observed in the interstellar \ion{Ca}{ii}K line. The values for AV\,304
have been taken directly from Rolleston et al. (\cite{rol03}).}
\label{t_CaII}
\centering
\begin{tabular}{lclc} \hline \hline
\multicolumn{2}{c}{NGC\,346-11} & \multicolumn{2}{c}{AV\,304} \\
Component & $v_{\rm lsr}$ (km~s$^{-1}$) & Component & $v_{\rm lsr}$ (km~s$^{-1}$) \\ \hline 
N 1  & 2    & A 1 & -5   \\
N 2  & 75   & A 2 & 13   \\
N 3  & 92   & A 3 & 44   \\
N 4  & 119  & A 4 & 64   \\
N 5  & 140  & A 5 & 92   \\
N 6  & 156  & A 6 & 110  \\
N 7  & 201  & A 7 & 122  \\
     &      & A 8 & 136  \\
     &      & A 9 & 157  \\
     &      & A 10& 171  \\
     &      & A 11& 187  \\
     &      & A 12& 200  \\
\hline
\end{tabular}
\end{table}

\section{Chemical composition of the SMC}       \label{s_SMC}

\begin{table}
\caption[]{Present-day chemical composition of the SMC as derived from
NGC\,346-11 and AV\,304. Also included are the results from some
recent studies of B-type stars, viz. Korn et al. (\cite{kor00})
and Trundle et al. (\cite{tru04}) and \ion{H}{ii} regions in the 
SMC, viz. Kurt et al. (\cite{kur99}); Reyes (\cite{rey99}); 
Testor (\cite{tes01}).}
\label{t_SMC}
\centering
\begin{tabular}{lcccccc} \hline \hline
 & \multicolumn{3}{c}{B-type stars}& \multicolumn{3}{c}{\ion{H}{ii} regions} \\ 
& This  & Korn   & Trundle & Kurt   & Reyes & Testor \\ 
& paper & et al. & et al.  &  et al. &       & \\ \hline
\ion{C}  & 7.42 & 7.40 & 7.30 & 7.53 & 7.39 & -    \\
\ion{N}  & 6.55 & 7.25 & 7.67 & 6.59 & 6.55 & 6.64 \\
\ion{O}  & 8.07 & 8.15 & 8.14 & 8.05 & 7.96 & 8.00 \\
\ion{Mg} & 6.77 & 6.78 & 6.78 & -    & -    & -    \\
\ion{Si} & 6.64 & 6.84 & 6.74 & 6.70 & -    & -    \\
\ion{Ne} & 7.84 & -    & -    & 7.26 & 7.17 & 7.23 \\
\ion{Al} & 5.33 & 5.58 & -    & -    & -    & -    \\
\ion{S}  & 6.40 & -    & -    & 6.42 & 6.32 & 6.31 \\
\ion{Fe} & 6.63 & 6.82 & -    & -    & -    & -    \\
\hline
\end{tabular}
\end{table}

Table~\ref{t_SMC} presents our best estimate for the present-day chemical 
composition of the SMC as derived from the photospheric abundances of 
NGC\,346-11 and AV\,304. In the following subsections we discuss in detail 
the methods used to derive these results from the individual abundances 
listed in Table~\ref{t_atmos+abund}.

\subsection{Carbon}                                              \label{s_C}

The carbon abundance was derived by taking the average of the abundance 
estimates from \ion{C}{ii} and \ion{C}{iii} in each star weighted by the 
inverse square of their estimated uncertainties. An alternative approach
would have been to only use the values showing the lowest uncertainities, viz. 
the \ion{C}{iii} abundance from NGC\,346-11 and the \ion{C}{ii} abundance from
AV\,304. As the abundance estimates agree relatively well,
it is not critical which method we adopt. The former method was used to 
maintain consistency with the rest of the analysis but using the latter 
resulted in an absolute abundance only 0.04 dex lower. 

\subsection{Nitrogen}                                              \label{s_N}

The nitrogen abundance estimated from the 3995\,\AA\ line in the spectra of
NGC\,346-11 should probably be treated as an upper limit as this weak line is 
poorly observed in our spectrum. For this reason we have adopted the
AV\,304 abundance of nitrogen as the nitrogen abundance in the SMC. 
If the NGC\,346-11 result had also been included,
the resultant estimate, considering the errors would have been only 0.05 dex higher
than the value listed in Table~\ref{t_SMC}. 

\subsection{Oxygen}                                                 \label{s_O}

Using the same method as discussed for carbon the SMC oxygen abundance was
derived. We note that the value deduced for the two stars differ by
0.3 dex and it is unclear whether this represents a real difference or
is due to uncertainties in, for example, the estimation of the atmospheric
parameters, the different sets of lines used for the two stars or the adopted
model ions. To investigate this we have undertaken differential analyses
with respect to two Galactic B-type stars, viz. HR\,2387 and $\tau$~Sco.
The former has atmospheric parameters ($T_{\rm eff}$ = 25500\,K; log $g$ =3.8;
$\xi$ = 5 km s$^{-1}$) similar to AV\,304, whilst those of the latter 
($T_{\rm eff}$ = 30500\,K; log $g$ =4.2; $\xi$ = 5 km s$^{-1}$) are similar 
to NGC\,346-11. The differential abundances are then -0.57$\pm$0.09 for 30
lines in AV\,304 and -0.70$\pm$0.09 for 16 lines in NGC\,346-11. Hence although
there remains a discrepancy, this is significantly reduced, indicating that
it probably does not represent a real difference in the stellar O abundances.

\subsection{Magnesium}                                              \label{s_Mg}

The derived magnesium abundance from each star was the same and the value has
been directly transferred into Table~\ref{t_SMC}.

\subsection{Silicon}                                                \label{s_Si}

Silicon is the only element where our choice of method for the combination of
the abundances derived from the individual species in each star becomes
important. The abundance estimates for AV\,304 are 0.34 dex higher for both 
\ion{Si}{iii} and \ion{Si}{iv} compared to the values derived for NGC\,346-11. 
If we adopt the same method to combine the results as was used for the 
other elements, a value of 6.60 dex would be obtained. However the errors 
associated with the estimates for NGC\,346-11 are dominated by the uncertainty 
in the adopted microturbulence particularly for the relatively strong \ion{Si}{iv}
lines. Indeed if a zero microturbulence was adopted, the silicon abundance 
estimate for NGC\,346-11 would increase to 6.60$\pm$0.15 dex. Additionally one 
could argue that the \ion{Si}{iv} lines in AV\,304 should not be used  
as these lines are sensitive to changes in the stellar parameters and 
especially the effective temperature. Hence we have opted to use only 
the \ion{Si}{iii} lines using the same weighting as for the other ions. We note
that this value is only 0.04 dex higher than that obtained by considering
both ionic species. We also note that a differential analysis similar to that
descibed in Sect. \ref{s_O} leads to values of -0.74$\pm$0.03 (from three lines)
for AV\,304 and -0.98$\pm$0.09 (from six lines) for NGC\,346-11, which as for 
oxygen leads to a decrease in 
the discrepancy between the two stars although it still remains significant.
It is unclear if this represents a real discrepancy or is an effect of the high
sensitivity of the Si lines to the stellar parameters. 

\subsection{Other elements}                                      \label{s_other}

Lines of other elements were only observed in the spectra 
of AV\,304, and the abundances in Table~\ref{t_SMC} come directly from 
those derived for this star and listed in Table~\ref{t_atmos+abund}. 
For neon, aluminium and iron, these should be treated with caution given
the limited number of features observed and the neglect of non-LTE effects.

Dufton et al. (\cite{duf86}) found that non-LTE effects in \ion{Al}{iii} were
relatively small and hence the abundance estimate for this species may
be reliable. For \ion{S}{iii}, there are a relatively large number of well
observed features and the uncertainty attached to the mean abundance is
relatively small, reflecting in part the good agreement between the
abundance estimates and hence this value 
should also be reliable. For both \ion{Ne}{ii} and \ion{Fe}{iii}, the agreement
between the estimates from individual lines is relatively poor and these
results may be unreliable with the iron value possibly being an underestimated, see 
Sect.~\ref{s_LTE}. Additionally, adopting this value leads to an iron abundance that
is approximately 0.3 dex lower than that used in the SMC grid. However as discussed
above, the use of models with a lower Fe abundance leads to similar results.


\section{Comparison with other SMC B-type stellar and \ion{H}{ii} region studies}      
                                                               \label{s_smc_HII}

In Table~\ref{t_SMC} we also present a comparison of our best estimate
for SMC abundances
with those from other studies of B-type stars and of \ion{H}{ii} regions. For
the former we have included the non-LTE analysis using static atmospheres
of Korn et al. (\cite{kor00}) and the non-LTE unified model atmosphere
analysis of Trundle et al. (\cite{tru04}). The three stars analysed by
Korn et al. (\cite{kor00}) have logarithmic gravities in the range 2.67 to 2.93 dex
and although described by Korn et al.(\cite{kor00}) as `non-supergiants', they have 
clearly evolved away from the main sequence. The targets of Trundle 
et al. (\cite{tru04}) are mostly Ia type supergiants
with gravities in the range 1.7 to 3.1 dex. There are numerous 
analyses of \ion{H}{ii} regions in the SMC including Russell \&
Dopita (\cite{rus92}), Reyes(\cite{rey99}), Garnett (\cite{gar99}), Kurt et al.
(\cite{kur99}), Peimbert et al. (\cite{pei00}) and Testor (\cite{tes01}).
Their abundance estimates are generally in quite good agreement and we 
have therefore listed three representative analyses in Table~\ref{t_SMC}.

Sofia and Meyer (\cite{sof01}) suggest that B-type stars may actually
underestimate the chemical composition of the ISM of the Galaxy and that young F
and G-type stars may better represent the ISM. Given the good agreement of our
stellar analysis with analyses of \ion{H}{ii} regions there is no evidence that
B-type stars underestimate the chemical composition of the ISM in the lower
metallicity environment of the SMC, although this possibility cannot be completely discounted.

\subsection{Carbon, Nitrogen and Oxygen}                   \label{s_CNO}

The carbon and oxygen abundance estimates for the three stellar studies 
are in good agreement, whilst the nitrogen estimates for the evolved 
stars are significantly higher than that found here. As discussed by both 
Korn et al.(\cite{kor00}) and Trundle et al. (\cite{tru04}), this probably 
reflects mixing of 
nucleosynthetic material to the surface of the evolved stars. Confirmation 
of this is provided by the excellent agreement between our nitrogen abundance 
estimate and that found in the \ion{H}{ii} regions studies. As the inital nitrogen abundance is much lower than those of
carbon or oxygen it only requires the processing of a relatively small amount of these elements into nitrogen to produce a significant
fractional increase in the nitrogen abundance in the atmospheres of evolved stars (Rolleston et al. \cite{rol03}). For example,
Trundle et al. (\cite{tru04}) and Dufton et al. (\cite{duf05}) have found nitrogen enhancements of approximately 1.0\,dex in SMC
supergiants. However, as discussed by these authors, the evolutionary models of Maeder \& Meynet (\cite{mae01}) would imply relatively
small corresponding C depletions (\textless\,0.25\,dex). This is consistent with the near normal C abundances found in SMC supergiants.

Our carbon and oxygen 
abundance estimates are also in good agreement with the \ion{H}{ii} region studies
and should probably be preferred over the other B-type studies as they are
less likely to be affected by small amounts of mixing of nucleosynthetic material to the
surface. Heap et al. (\cite{hea04}) have derived abundances from 17 O-type stars
in the SMC and these are in excellent agreement with our abundance estimates for cabon
and oxygen. From their three targets that do not appear to show enhanced nitrogen they
derive a nitrogen abundance 0.15\,dex lower than that given here. This may not
be significant given the errors in our estimate and the weakness of the
\ion{N}{ii} spectrum in our targets.

\subsection{Magnesium and Silicon}                          \label{s_MgSi}

All the B-type stellar studies yield very similar Mg values probably 
reflecting the relatively simple term structure for this ion and 
small non-LTE effects found for B-type stars (see, for example, 
Mihalas \cite{mih72}). No comparison was possible with the
\ion{H}{ii} region studies but our estimate is
in excellent agreement with that deduced for A-type 
(Venn \cite{ven99}; 6.82\,dex) and for late-type (see, for example,
Spite et al. \cite{spi89}, \cite{spi91}; 6.79\,dex) stars.
The silicon abundance derived here should be treated with some caution given
the difference in our estimates for our two targets. Nevertheless, our 
reported value agrees well (within our uncertainties) with the B-type supergiant 
value of Trundle et al. and the \ion{H}{ii} region value of Kurt et al. It is
also in agreement with the late-type stellar value of Spite et al. 
(\cite{spi89}, \cite{spi91}; 6.75\,dex) but is lower than the B-type estimate of
Korn et al. and the A-type estimate of Venn (\cite{ven99}; 6.97\,dex). Our silicon
abundance is also in reasonable agreement with the value derived from O-type stars
(Heap et al. \cite{hea04}; 6.85\,dex).

\subsection{Neon, Aluminium, Sulphur and Iron}            \label{s_NeAlSFe}

As discussed above the estimates for neon, aluminium and iron are based on LTE
calculations and hence may be less reliable. For aluminium and iron, our 
estimates are in reasonable agreement with the B-type stellar study of
Korn et al. (\cite{kor00}), although  we find a relatively 
poor agreement between the values deduced from individual features of iron.
Additionally, the B-type estimates for iron are lower than
those found in A-type (Venn \cite{ven99}; 6.70 dex) and late-type 
(Spite et al. \cite{spi89}, \cite{spi91}; 6.72 dex) stars. Peters \& Adelman
(\cite{pet02}) have derived an iron abundance for AV\,304 from Far Ultraviloet
Spectroscopic  Explorer (FUSE) data which is 0.2\,dex higher than our value
although considering the uncertainity in the estimate this may not be
significant. We are also able to compare our iron abundance to that derived from
O\,stars by Bouret et al. (\cite{bou03}) and Heap et al. (\cite{hea04}) and our 
estimate is again lower than the values derived therein. It is
therefore possible that our iron abundance is underestimated due to
\ion{Fe}{iii} not being the dominant ionization stage as discussed in
Sect.~\ref{s_LTE}. 

For neon, our estimate is 
significantly larger than those found from \ion{H}{ii} region studies. Given
the spread of values estimated from the three stellar features and the
lack of any non-LTE calculations for this element, the latter should 
be considered more reliable. By contrast our sulphur abundance
estimate is based on a relatively large number of features that show
reasonable internal agreement. Additionally, it is in excellent agreement
with the \ion{H}{ii} region studies and as it has been treated using non-LTE
methods it should be considered as reliable, although it is lower than the O-type stellar results of 
Heap et al.(\cite{hea04}).

\section{Conclusions}                                         \label{s_concl}

The principle aim of this analysis was to provide abundance
estimates for two unevolved B-type stars in the SMC. These should be
representative of their progenitor interstellar material and as such
should complement those deduced from \ion{H}{ii} region studies.
Additionally the high quality of the observational material (coupled
with the small stellar projected rotational velocities) and the
use of non-LTE model atmosphere techniques should lead to more reliable
estimates than have previously been deduced from main sequence SMC
targets. For the ions included in the non-LTE calculations,  
our estimates for C, N, O, Mg and S should be reliable with that for
Si being slightly more uncertain. It is difficult to quantify
the uncertainties in these values and in particular those arising from 
the physical assumptions adopted. However from our estimated uncertainties
and from a comparison with other studies, we believe that these estimates 
should normally be accurate to $\pm$0.2\,dex or better. For the ions
considered in an LTE approximation, the estimates for Al may
be reliable, whilst those for Ne and Fe should be treated with caution.

Assuming that the SMC is well mixed, these estimates provide the best
estimates of the chemical composition of the SMC as available from
unevolved B-type stars. As such they will be particularly useful in
interpreting the results from studies of other, possibly evolved, 
B-type stars, which have utilised similar methods.

\begin{acknowledgements}
We are grateful to the staff of the European Southern Observatory for help 
obtaining the observational data. RSIR and WRJR acknowledges financial 
support from PPARC (grant no G/O/2001/00173), DJL acknowledges support 
through QUB's visiting fellow programme (grant no V/O/2000/00479).
\end{acknowledgements}


\end{document}